\documentclass[10pt,aps,prb,twocolumn,showpacs,amsmath,amssymb,superscriptaddress,floatfix]{revtex4-1}
\usepackage{graphicx}
\usepackage{dcolumn}
\usepackage{bm}
\usepackage{epsfig}
\usepackage{multirow}
\usepackage{rotating}
\usepackage{xcolor}
\graphicspath{{../combined-figures/}}

\begin{document}
\title{Quantum magnetism in molecular spin ladders probed with muon-spin spectroscopy}
\author{T. Lancaster}
\email{tom.lancaster@durham.ac.uk}
\affiliation{Centre for Materials Physics, Durham University, Durham,
  DH1 3LE, United Kingdom}
\author{F. Xiao}
\affiliation{Centre for Materials Physics, Durham University, Durham,
  DH1 3LE, United Kingdom}
\affiliation{Laboratory for Neutron Scattering and Imaging, Paul Scherrer Institut, CH-5232 Villigen PSI, Switzerland}
\affiliation{Department of Chemistry and Biochemistry, University of Bern, CH-3012 Bern, Switzerland}
\author{B.M. Huddart}
\affiliation{Centre for Materials Physics, Durham University, Durham,
  DH1 3LE, United Kingdom}
\author{R.C. Williams}
\affiliation{Centre for Materials Physics, Durham University, Durham,
  DH1 3LE, United Kingdom}
\author{F.L. Pratt}
\email{francis.pratt@stfc.ac.uk}
\affiliation{ISIS Facility, STFC Rutherford Appleton Laboratory,
  Chilton, Didcot, Oxfordshire, OX11~0QX, United Kingdom}
\author{S.J. Blundell}
\affiliation{Oxford University Department of Physics, Clarendon Laboratory, Parks
Road, Oxford, OX1~3PU, United Kingdom}
\author{S.J. Clark}
\affiliation{Centre for Materials Physics, Durham University, Durham,
  DH1 3LE, United Kingdom}
\author{R.  Scheuermann}
\affiliation{Laboratory for Muon Spin Spectroscopy, Paul Scherrer Institut, CH-5232 Villigen PSI, Switzerland}
\author{T. Goko}
\affiliation{Laboratory for Muon Spin Spectroscopy, Paul Scherrer Institut, CH-5232 Villigen PSI, Switzerland}
\author{S. Ward}
\affiliation{Laboratory for Neutron Scattering and Imaging, Paul Scherrer Institut, CH-5232 Villigen PSI, Switzerland}
\author{J.L. Manson}
\affiliation{Eastern Washington University, Department of Chemistry and Biochemistry, Cheney, Washington 99004, USA}
\author{Ch.\ R\"{u}egg}
\affiliation{Laboratory for Neutron Scattering and Imaging, Paul Scherrer Institut, CH-5232 Villigen PSI, Switzerland}
\affiliation{Department of Quantum Matter Physics, University of Geneva, CH-1211 Geneva, Switzerland}
\author{K.W. Kr\"{a}mer}
\affiliation{Department of Chemistry and Biochemistry, University of Bern, CH-3012 Bern, Switzerland}
\date{\today}

\begin{abstract}
We present the results of muon-spin spectroscopy ($\mu^{+}$SR) measurements on the molecular spin ladder system (Hpip)$_{2}$CuBr$_{4(1-x)}$Cl$_{4x}$, [Hpip=(C$_{5}$H$_{12}$N)].
 Using transverse field $\mu^{+}$SR we are able to identify characteristic behaviour in each of the regions of the phase diagram of the $x=0$ strong-rung spin ladder system (Hpip)$_{2}$CuBr$_4$. Comparison of our results to those of the dimer-based molecular magnet Cu(pyz)(gly)(ClO$_{4}$) shows several common features.
We locate the crossovers in partially disordered (Hpip)$_{2}$CuBr$_{4(1-x)}$Cl$_{4x}$ ($x=0.05$), where a region of behaviour intermediate between quantum disordered and Luttinger liquid-like is identified. 
Our interpretation of the results incorporates an analysis of the probable muon stopping states in (Hpip)$_{2}$CuBr$_4$  based on density functional calculations and suggests how the muon plus its local distortion can lead to a local probe unit with good sensitivity to the magnetic state.
Using longitudinal field $\mu^{+}$SR we compare the dynamic response of the $x=1$
 strong-rung material (Hpip)$_{2}$CuCl$_{4}$  to that of the
 strong-leg material (C$_{7}$H$_{10}$N)$_{2}$CuBr$_{4}$ (known as DIMPY) and demonstrate that our results are in agreement with predictions based on interacting fermionic quasiparticle excitations in these materials. 
\end{abstract}
\maketitle

\section{Introduction}
\begin{figure}
\begin{center}
\epsfig{file=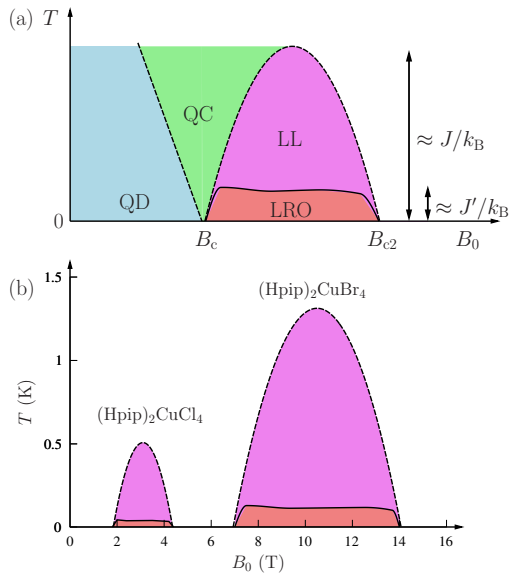, width=7cm}
\caption{(a) Schematic temperature-magnetic field phase diagram for
 strong-rung spin ladder compounds. The
coherence scale of the order of the exchange parameter $J$
along the legs of the ladder is shown. Weak interladder
coupling $J'$ leads to LRO. (b) Schematic LL domes for (Hpip)$_2$Cu$X_4$, $X=\ $Br and Cl, with the dome height determined via thermodynamic measurements. 
\label{fig:phasediagram}}
\end{center}
\end{figure}

Spin ladders represent a class of low-dimensional quantum magnets that occupy a 
 regime of subtle  which lies between the stark extremes of the one-dimensional chain or two-dimensional plane \cite{giamarchi}. 
These materials are
 characterised by two antiferromagnetic exchange parameters: $J_{\mathrm{rung}}$ along the
 ladder rungs and $J_{\mathrm{leg}}$ along the ladder legs. For ladders with an even number of legs the ground states are magnetically disordered and show a gap in their excitation spectrum. [They are often described as quantum disordered (QD)]. An applied magnetic field $B_{0}$ acts to close the gap and, at a critical field $B_{0}=B_{\mathrm{c}}$, there exists a $T = 0$ quantum critical point (QCP) above which the excitation spectrum is gapless. In a one-dimensional spin system such as an isolated ladder, divergent phase fluctuations prevent the possibility of the high-field, gapless state showing long range magnetic order (LRO) and, instead, a spin Luttinger 
liquid (LL) state is realised. This gapless LL state is characterised by algebraically decaying spin correlations and, since it involves no symmetry breaking, is reached via a crossover rather than a phase transition for $T>0$. Despite this, the change in spin correlations 
has been shown to be observable in thermodynamic measurements \cite{ruegg,thielemann}. 
The QCP and related physics in two-leg ladders has been extensively studied theoretically, but there has been comparatively little matching experimental work due to the 
scarcity of model systems with accessible energy scales.  
Fig.~\ref{fig:phasediagram}(a) shows a schematic applied magnetic field-temperature phase diagram 
for a spin ladder.

Arguably the best studied and clearest example of LL physics has been found in a two-leg spin ladder system: the strong-rung coordination compound piperidinium copper halide (Hpip)$_2$Cu$X_4$[where (Hpip) = (C$_5$H$_{12}$N) and $X =\ $Br, Cl], whose phase diagrams are shown in Fig.~\ref{fig:phasediagram}(b). 
The $X=\ $Br compound (Hpip)$_2$CuBr$_{4}$ has\cite{ruegg,thielemann,klanjsek} $J_{\mathrm{leg}}/J_{\mathrm{rung}}= 0.25$ and a phase diagram that reveals LL, quantum critical (QC) and QD regimes at low temperature, with a critical field, derived from magnetization measurements, of $B_{\mathrm{c}} = 6.7$~T at $T=0$. Owing to the existence of a small three-dimensional coupling $J'$ there is also a region of 3D LRO within the LL dome below $T \approx 0.1$~K, where the inter-ladder exchange coupling leads to 3D magnetic  reminiscent of that shown by coupled spin dimers in an applied magnetic field \cite{lancaster}. Above a saturation field $B_{\mathrm{c}2}$ the system becomes gapped, once again.  
The $X=\ $Cl compound is characterised\cite{ward,ryll} by $J_{\rm leg}/J_{\rm rung} \approx 0.39$ and  a critical field of $B_{\mathrm{c}} = 1.73$~T.
Taking these two compounds as 
end points in a series (Hpip)$_{2}$CuBr$_{4(1-x)}$Cl$_{4x}$, the controlled introduction of disorder via halide substitution $x$, 
offers opportunities for studying the effects of disorder in LL physics \cite{ward}. 
The substitution of the Br and Cl halide ions, which mediate the superexchange interactions between the Cu$^{2+}$ spin sites, facilitates disorder via randomised interaction strengths whilst
leaving the spin ladder structure qualitatively unchanged. This quenched disorder provides the possibility of realising and characterising new quantum states, such as the elusive Bose glass phase \cite{ward,zheludev}. 
A summary of the parameters of representative members of the family of materials discussed in this paper is given in Table~\ref{table1}. 
\begin{table}
\begin{tabular}{cccccc}
\hline
Compound & $J_{\mathrm{rung}}$(K) & $J_{\mathrm{leg}}$ (K) & $B_{\mathrm{c}}$ (T) & $B_{\mathrm{c2}}$ (T)\\ 
\hline
(Hpip)$_{2}$CuBr$_{4}$\cite{klanjsek} & 12.6 &3.55 &6.73 &13.79\\
(Hpip)$_{2}$CuBr$_{2}$Cl$_{2}$\cite{tajiri}& 5.10 &3.06 &2.4 &20\\
(Hpip)$_{2}$CuCl$_{4}$ \cite{ryll}& 3.42 & 1.34 &1.73 &4.38\\
DIMPY\cite{jeong} & 9.5 & 16.5 & 3.0 & 29\\
\hline
\end{tabular}
\caption{Summary of ladder exchange parameters and critical fields
from magnetization data. In addition to the materials measured in this study, also included is the half-substituted (Hpip)$_{2}$CuBr$_{2}$Cl$_{2}$. \label{table1}}
\end{table}

 In this paper we present local probe measurements of the magnetism in strong rung spin ladders. We employ implanted muons\cite{yaouanc,steve} in two configurations. 
Firstly, 
the transverse field (TF) geometry, intended to probe the static magnetic field distribution in (Hpip)$_{2}$CuBr$_{4}$ and in partially disordered (Hpip)$_{2}$CuBr$_{4(1-x)}$Cl$_{4x}$.
Secondly, the longitudinal field (LF)  geometry, intended to probe dynamics and applied to 
 (Hpip)$_2$CuCl$_{4}$.
Muon spin relaxation ($\mu^{+}$SR) has previously been shown to be a sensitive local probe of static and dynamic effects in one- and two-dimensional coordination polymer molecular magnets \cite{tom1,steele,fan}, in both TF and LF configurations. 
The technique has also been shown to usefully probe the phase diagram  and excitations in systems based on coupled antiferromagnetic dimers\cite{lancaster} and  in spin liquid systems\cite{pratt},  whose low energy physics can also be viewed as being related to that of antiferromagnetically coupled, strongly interacting dimers. 
However, detailed
investigations of spin ladders using $\mu^{+}$SR have not been possible until
the recent commissioning of new
spectrometers with high magnetic field and low temperature capabilities \cite{hifi,hal}.

There is also a direct and useful correspondence between a spin Hamiltonian in an applied magnetic field and a lattice boson Hamiltonian in the grand canonical ensemble \cite{zheludev}, where the applied magnetic field acts as an effective chemical potential for the boson excitations. 
For the 3D interacting dimer case, this gives us a picture of the magnetically ordered regime as being
an analogue of a Bose Einstein condensate (BEC) for magnons.
For the 1D spin ladder case, this picture shows that the applied field provides us with a control over the population of excitations within the LL state.
The measurements on spin ladders presented here may therefore be viewed as providing a bridge between local probe results on the relatively  well understood 3D dimer physics seen in systems such as Cu(pyz)(gly)(ClO$_{4}$) \cite{lancaster} and the BEC of magnons candidate NiCl$_{2}\cdot$4SC(NH$_{2}$)$_{2}$ \cite{sjb}, as well as the more exotic case of spin liquid physics examined, for example, in Ref.~\onlinecite{pratt}.
Since spin ladder systems lie between these extremes, an understanding
of the muon's interaction with spin ladders is useful in interpreting
the results of the more speculative work in the more complex spin liquid systems. Moreover, in cases such as the spin liquids, where $\mu^{+}$SR has provided unique insights, it is especially important to assess the extent to which the implanted muon has the potential to perturb the intrinsic magnetic state. 

We show here the ways in which  implanted muons are sensitive to the different phases in the spin ladder materials and also to low frequency dynamics. Notably, our first principles calculations of the nature of the muon stopping site suggest that the muon's sensitivity to the physics in these systems derives from a local distortion it makes to the crystal structure in its vicinity, which leads to a significant perturbation to the local magnetism
(though in a quite different manner to that found in pyrochlore oxides \cite{foronda}). Despite this, the muon continues to prove a useful probe of the global magnetic properties of the materials, allowing us in the case of (Hpip)$_{2}$CuBr$_{4}$, for example,  to identify phase boundaries in agreement with those suggested by other techniques.

The paper is structured as follows: Section~\ref{expt} contains a brief description of the experimental geometries employed in this investigation. 
In Section~\ref{tf} we present the results of TF $\mu^{+}$SR on (Hpip)$_{2}$CuBr$_{4}$
and its comparison with measurements made on the dimer-based molecular magnet Cu(pyz)(gly)(ClO$_{4}$), where similar spectra are measured.
We analyze in detail the state of the stopped muon in Section~\ref{muonsites} using first principles techniques and suggest how muon-induced changes in the local electronic structure lead to a muon state that is sensitive to the local magnetism. 
In Section~\ref{tf2}, we use the sensitivity of the muon to the local magnetism to probe the  effect of introducing bond disorder.
Finally, in Section~\ref{lf} we turn directly to dynamics and discuss the comparison of LF $\mu^{+}$SR measurements made on strong-rung (Hpip)$_{2}$CuCl$_{4}$ and strong-leg DIMPY.  

\section{Experiment}\label{expt}

In the TF $\mu^+$SR experimental geometry \cite{steve} the externally  applied field  $B_{0}$ is directed perpendicular to the initial muon spin direction. Muons precess about the total local magnetic field $B$ at the muon site. The observed property of the experiment is the time evolution of the muon spin polarization $P_x(t)$, which allows the determination of the distribution $p(B)$ of local magnetic fields across the sample volume by means of a Fourier transform.
TF $\mu^+$SR measurements on 
(Hpip)$_{2}$CuBr$_{4(1-x)}$Cl$_{4x}$ ($x=0, 0.05$) were carried out on single crystal samples at the Swiss Muon Source, Paul Scherrer Institute (Switzerland) using the HAL-9500 high field spectrometer.
 The crystals had approximate dimensions $5\times 5\times 1$~mm$^{3}$.
These cover the muon  beam area for the HAL-9500 spectrometer, leading us to expect very little background signal from muons stopping in the sample holder. 
The single crystal samples were wrapped in Ag foil (thickness 12~$\mu$m) and glued to a silver sample holder with GE varnish. The holder was
mounted on the cold finger of a dilution refrigerator,  again with the field directed along the $b$-axis. 
Analogous measurements were made on a polycrystalline mosaic of crystallites of Cu(pyz)(gly)(ClO$_{4}$) using HAL-9500. The crystals of this material were arranged on an Ag foil (thickness 12.5~$\mu$m) which was glued to the sample holder. 
Data analysis was carried out using the WiMDA analysis program\cite{wimda}, with TF spectra generated using WiMDA's apodized, phase-corrected cosine Fourier transforms.

In the LF $\mu^{+}$SR experimental geometry the field $B_{0}$ is directed parallel to the initial muon-spin polarization. The applied field decouples the contribution
from static magnetic fields at the muon site. This allows us to
probe the dynamics of the system, as time-varying magnetic
fields at the muon site are able to flip muon spins and therefore to relax the average muon polarization. Our LF measurements were made using the HiFi spectrometer at the STFC-ISIS facility, Rutherford Appleton Laboratory (UK). A single crystal sample of (Hpip)$_{2}$CuCl$_{4}$ was mounted on an Ag plate attached to the cold finger of a dilution refrigerator, with the field directed along the $b$-axis  (i.e.\ perpendicular to the ladders). The sample sizes were similar to those given above.

\section{Transverse field measurements}\label{tf}

\subsection{(Hpip)$_{2}$CuBr$_{4}$}

\begin{figure}
\begin{center}
\epsfig{file=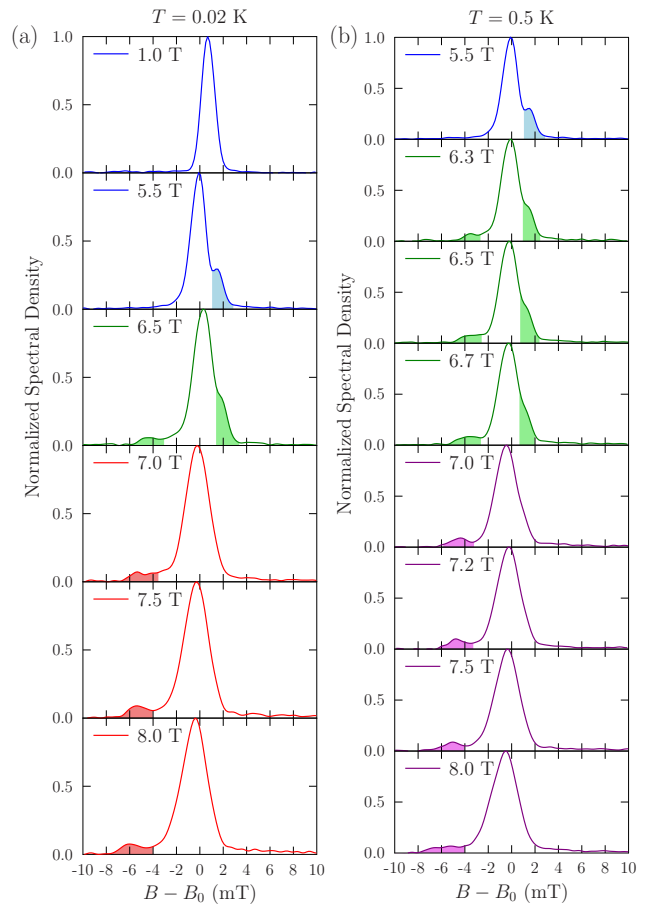, width=\columnwidth}
\caption{Fourier transform TF $\mu^{+}$SR spectra measured for (Hpip)$_{2}$CuBr$_{4}$ at fixed temperatures of  (a) $T=0.02$~K and (b) $T=0.5$~K. Colors refer to Fig.~1(a) and shading highlights the satellite peaks. \label{fig:spectra}}
\end{center}
\end{figure}

\begin{figure}
\begin{center}
\epsfig{file=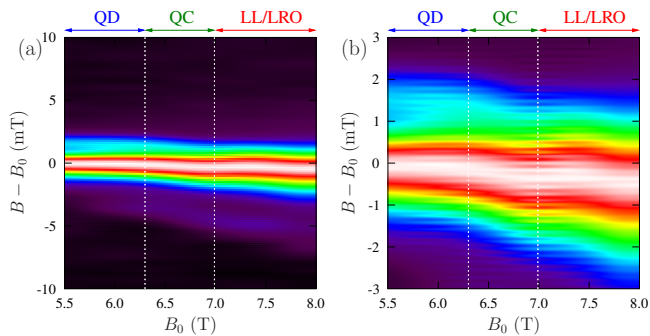, width=\columnwidth}
\caption{A color map of Fourier amplitude as a function of applied field $B_{0}$ for TF measurements on (Hpip)$_{2}$CuBr$_{4}$.
Data are shown at 
$T=0.5$~K, with a broad vertical scale on the left and an expanded vertical scale on the right. 
White lines are guides to the eye showing probable transitions (see text). \label{fig:contours}}
\end{center}
\end{figure}

To investigate the phase diagram of the spin ladder system using implanted muons, we use TF $\mu^{+}$SR measurements. In the high magnetic field limit, the muon spin relaxation rate (and hence the width of the features seen in the Fourier transforms of the spectra) are determined by 
the magnetic field correlations along the direction of the applied magnetic field. 
Example Fourier transform TF $\mu^+$SR spectra are shown for (Hpip)$_{2}$CuBr$_{4}$ in Fig.~\ref{fig:spectra} at two temperatures. At each applied field we find significant Fourier amplitude $A(B)$ [proportional to the field distribution $p(B)$] close to the applied field $B_{0}$, but also significant spectral weight displaced from $B_0$. With increasing applied field the average spectral weight shifts to lower fields. 
The evolution of the spectral features may also be tracked in the color map plot of Fourier amplitude shown in 
Fig.~\ref{fig:contours}.

\begin{figure}
\begin{center}
\epsfig{file=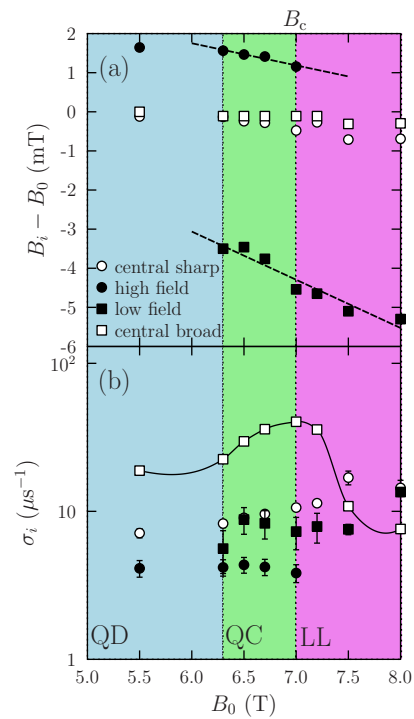, width=6.cm}
\caption{Results of fitting Gaussian peaks in Eq.~\ref{fitting1} to the data for (Hpip)$_{2}$CuBr$_{4}$ at $T=0.5$~K. (a) Peak position $B_{i}$ relative to the applied field $B_{0}$. (Lines are guides to the eye.)  (b) Peak width $\sigma_{i}$. \label{fig:fitting-undoped}}
\end{center}
\end{figure}

In order to extract more quantitative detail from the field distributions,  the spectral  functions $A(B)$ were fitted to the sum of several peaks. It is found that for $5.5\leq B_{0} \leq 8$~T the data can be best described by a sum of four distinct Gaussian components: two modelling the shape of the large feature whose centre is near $B_{0}$, one for the high field side feature seen in the QD region and one for the feature seen on the low-field side of the peak that persists into the LL regime. The spectra were therefore fitted to the function
\begin{equation}
A(B) = \sum_{i=1}^{4} 
A_{i} \exp\left[-\frac{(B-B_{i})^{2}} {2\sigma_{i}^{2}}\right].
\label{fitting1}
\end{equation}
The amplitudes $A_{i}$ were found to be roughly constant with varying applied field, allowing us to track the behaviour of each of the components.\cite{note1}  [We find that the two central components have similar spectra weight (given by the product $A_{i}\sigma_{i}$), with the peaks above and below each having $\approx$15\% of that spectral weight.]
The position $B_{i}$ and width $\sigma_{i}$ of the peak components are extracted and plotted against applied field $B_{0}$ in Fig.~\ref{fig:fitting-undoped} for data measured at $T=0.5$~K. 

Several trends are apparent in the data and the corresponding fits.
 As the quantum critical (QC) region is approached from the quantum disordered (QD) regime on increasing $B_{0}$ a high field shoulder is seen in addition to the large central peak. 
As the field is increased, this feature shifts towards the central line, before merging into the broad central peak above the crossover field $B_{\mathrm{c}} \approx 7$~T. At $B_{0}=6.3$~T and above, we see an additional peak on the low field side, which persists and shifts to still lower fields as the applied field is increased. The broader of the two central features shows a maximum in its width centred on $B_{\mathrm{c}}$. 
It is notable that these features correlate with the independently determined phase diagram of the material. 
The high field feature is visible in the QD and QC regimes and disappears as we pass into the LL/LRO regions. 
As we leave the QD region and cross into the QC regime, a low-field feature appears which persists into the LL/LRO region. 
Taking the $T=0$ QCP to be at the centre of the QC region leads to an estimate $B_{c}=6.7$~T, in broad agreement with the magnetization result.
  We are therefore able to broadly distinguish the regimes showing quantum disorder (with the high-field feature resolvable) and LL physics (where this feature is not resolved).


The spectra are generally quite similar for $T=0.02$~K and $T=0.5$~K where they reflect the same phases. The most significant differences are seen in the data measured for (Hpip)$_{2}$CuBr$_{4}$ around 6.5~T and 8 T. The former field is close to the QD--QC
crossover [Figs.~\ref{fig:phasediagram}(a) and (b)]. At this applied field, raising the temperature causes a shift in spectral weight to lower fields. This is consistent with the negative slope of the  boundary between QD and QC regions [Fig.~\ref{fig:phasediagram}(a)], which causes the low temperature point to be closer to the QD region and the higher temperature point to be closer to the QC regime. The shift in spectral weight to low fields is then seen to follow the same trend as we find in the  constant temperature scans from the QD to QC region, which also involve a shift in spectral weight to lower fields as the applied field is increased. 
At 8 T the 0.5~K data reflects the LL phase, with a rather broad low field satellite, whereas at 0.02~K the satellite found to be much sharper, reflecting the LRO phase. The overall difference between the LL and LRO phases is however found to be quite subtle.

\subsection{Cu(pyz)(gly)(ClO$_{4}$)}

\begin{figure}

\begin{center}
\epsfig{file=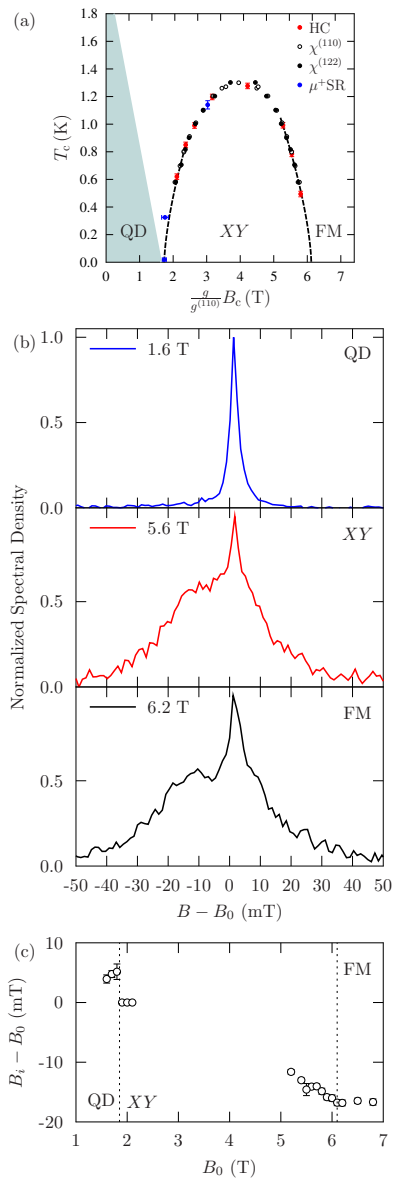, width=5.5cm}
\caption{(a) Phase diagram for Cu(pyz)(gly)(ClO$_{4}$) taken from Ref.~\onlinecite{lancaster} showing quantum disordered, $XY$-ordered and ferromagnetic (FM) regions. Phase boundaries are derived from heat capacity (HC) dynamic susceptibility ($\chi$) and $\mu^{+}$SR. 
(b) TF $\mu^{+}$SR spectra in the three phases of the material measured at $T=0.2$~K. 
(c) Separation of a fitted Lorentzian peak from the applied magnetic field as a function of applied field. Phase boundaries are indicated with dotted lines. 
\label{fig:glydata}}
\end{center}
\end{figure}

The features identified above reflecting the physics of the system can be compared with analogous measurements we have made on the magnetic dimer material Cu(pyz)(gly)(ClO$_{4}$). This system\cite{lancaster,brambleby} is based on 
interacting antiferromagnetic dimers which are coupled in a staggered configuration into sheets. The measured phase diagram of this system is shown in Fig.~\ref{fig:glydata}(a), showing quantum disordered, $XY$ antiferromagnetically ordered and ferromagnetically ordered (FM) phases as a function of applied field $B_{0}$. Our TF $\mu^{+}$SR spectra measured as a function of applied field $B_{0}$ [Fig.~\ref{fig:glydata}(b)] show the same trends as the spin ladder measurements. In the QD phase at 1.6~T the observed line is asymmetrical with spectral weight on the high field side of the central peak. In the ordered phase (5.6~T) the spectrum is significantly broadened with spectral weight shifted to the low field side of the spectrum. We fit Lorentzian peaks to these data using the function 
\begin{equation}
A(B) = \sum_{i=1}^{N} \frac{A_{i}}{\left[1+\left(\frac{B-B_{i}}{\sigma_{i}}\right)^{2}\right]},
\label{fit2}
\end{equation}
with $N=3$. 
This
allows us to track the separation $B_{i} -B_{0}$ of the additional spectral weight from the peak at the applied field. The Lorentzian lineshape most likely reflects a particularly strong dynamics in this case. The result of this analysis is shown in Fig.~\ref{fig:glydata}(c), where we see that the low-field phase boundary can be identified via a discontinuity in
$B_{i} -B_{0}$ with increasing $B_{0}$, while the high-field boundary may be identified via a discontinuity in the gradient of this quantity. 
The former feature is probably due to the change in both time-averaged local field and dynamics on crossing the QD-$XY$ phase boundary. 
The latter feature probably reflects the fact that although the $XY$-FM transition does involve a crossing of energy levels, so is indeed a phase transition, at least classically these two phases result in muon ensembles experiencing  similar local magnetic field profiles. 

This same general trend is also the case in measurements we have made 
\cite{sjb} on the BEC candidate material NiCl$_{2}\cdot 4$SC(NH$_{2}$)$_{2}$ (known as DTN). In that system, measurements in the QD regime show a peak on the high-field side of the applied field, which vanishes as the LRO phase is approached, with spectral weight appearing on the low-field side in the LRO region, as we observe here. 
It is interesting to note that a significant difference between the results for the spin ladder material and those for DTN and Cu(pyz)(gly)(ClO$_{4}$). For (Hpip)$_{2}$CuBr$_{4}$ we start to see the low-field feature emerge in the QC region,  which coexists with the high field peak.  In contrast there is a sharp separation in behaviour in DTN and Cu(pyz)(gly)(ClO$_{4}$) between the QD and LRO regions, with no coexistence.  This reflects the difference in physics between the two magnetic systems in which the lack of symmetry breaking in the spin ladder material leads to $T>0$ crossovers rather than the phase transitions of the BEC materials.

\section{The state of the stopped muon}\label{muonsites}

\subsection{Dynamical Regime}

In order to understand the origin of the TF $\mu^{+}$SR signals reported in section \ref{tf} and the LF $\mu^{+}$SR signals that will be reported in section \ref{lf} we examine the nature of the stopping state of the muon. 
We consider first the case of ZF and LF $\mu^{+}$SR. In most  magnets, we are well within the fast fluctuation limit at temperatures above $T_{\mathrm{N}}$. In zero field measurements we often observe that the electronic fluctuations are partially narrowed from the spectrum (allowing nuclear moments to make a sizeable contribution to the relaxation in zero field).
A rough estimate can be made of the muon relaxation rate based on the fluctuating amplitude of a magnetic field component at the muon site and the fluctuation time $\tau$.
In a fast fluctuation regime with dense magnetic moments and relaxation dominated by exponential correlations \cite{yaouanc} 
 we expect the muon polarization $P(t)$ to relax following an exponential function $P(t) = \exp(-\lambda_{\mathrm{LF}} t)$
with the LF relaxation rate given by
\begin{equation}
\lambda_{\mathrm{LF}} = \frac{2\gamma^{2} \Delta B^{2} \tau }{\omega_{0}^{2}\tau^{2}+1},
\end{equation}
where $\omega_{0} = \gamma B_{0}$, $B_0$ being the applied field and $\gamma$ the muon gyromagnetic ratio. 
If there are several distinct muon sites in a material, then we would expect a relaxation rate $\lambda_{i}$ for a particular site $i$, to reflect the fluctuating magnetic field $\Delta B_{i}$ at that site, along with the correlation time for the site $\tau_{i}$. 
The relaxation of the muon site would then follow $P(t) = \sum_{i} a_{i}\exp(-\lambda_{i}t)$, where $a_{i}$ reflects the occupancy of each site. 

Within the same approximation, the TF relaxation is given by the Abragam function\cite{yaouanc}, which in the fast fluctuation limit also predicts exponential relaxation with a transverse relaxation rate
\begin{equation}
\lambda_{\mathrm{TF}} = \gamma^{2}\Delta B^{2} \tau 
\end{equation}
and a corresponding Lorentzian lineshape in the frequency domain. Assuming that, for a particular site, transverse and longitudinal field-field correlations are of the same order of magnitude, we may compare the transverse relaxation rate  of $\approx 10$~$\mu$s$^{-1}$ with a typical  longitudinal one of $\approx 0.1$~$\mu$s$^{-1}$ (see section \ref{lf}), from which we obtain an estimate of the characteristic parameters $\Delta B \approx 50$~mT and $\tau \approx 5$~ns (corresponding to $\gamma\Delta B \tau \approx 0.2$, confirming the original assumption of being in the fast fluctuation limit). This would imply that there is at least one muon site experiencing significant dynamic fluctuations of this amplitude. Dipole field calculations based on the muon site analysis below (see Appendix) are fully consistent with this estimate. 

Besides this evidence for a fast fluctuating dynamical response in at least part of the spectrum, the TF measurements are also sensitive to any time averaged field component along the applied field direction. Evidence for such quasistatic fields is provided by the shifted satellite features found in the spectra.  

\subsection{Muon sites}

To identify the specific classes of muon stopping state we have carried out spin-polarized density functional theory (DFT) calculations within the generalized gradient approximation (GGA) \cite{PBE}, using the plane wave basis-set electronic structure code \textsc{Castep} \cite{CASTEP}.   Structural relaxations were performed using a $2 \times 1 \times 1$ supercell and details of these calculations can be found in the Appendix.

\begin{figure}[ht]
	\includegraphics[width=\columnwidth ]{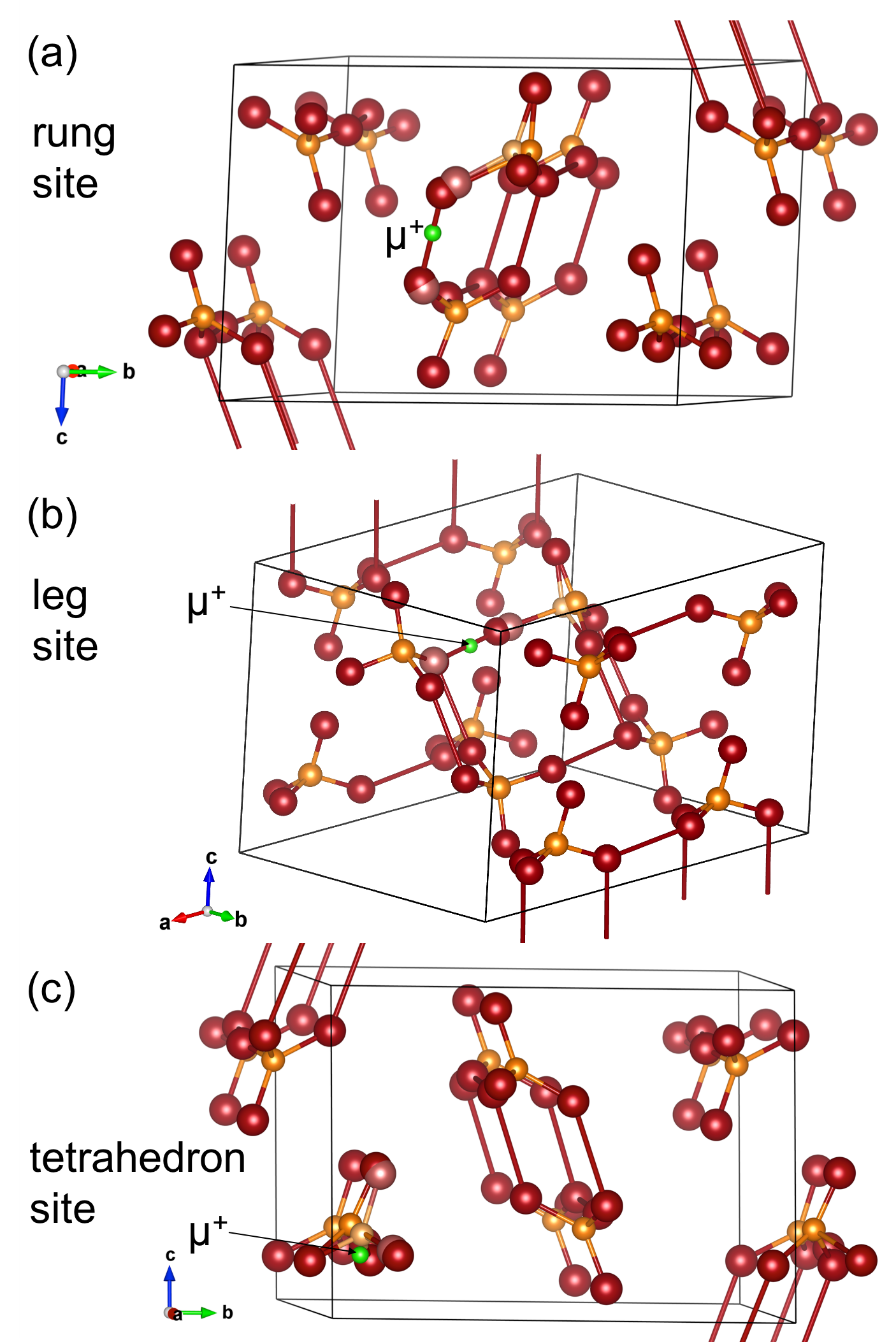}
	\caption{Muon stopping sites within the spin ladder structure.  Muons were found to stop (a) along ladder rungs (b) along ladder legs and (c) inside the CuBr$_4$ tetrahedra. For clarity, only Cu atoms (orange spheres) and Br atoms (red spheres) are shown.}
	\label{sites}
\end{figure}

We find three characteristic stopping sites based on the local geometry around the muon position and show each of these in Fig.~\ref{sites}.  (I) In the lowest energy sites the muon sits between two Br atoms forming a rung of the spin ladder [Fig.~\ref{sites}(a)] (called the {\it rung sites} below).   The resulting Br-$\mu^{+}$-Br -like structure is similar to the F-$\mu^{+}$-F complex\cite{brewer} commonly formed in other complex systems\cite{lancaster2} (and predicted using DFT \cite{moeller}), although we note in the present case that the two $\mu^{+}$-Br bonds lengths are unequal.  (II) A second class of site has the muon sitting between two Br atoms forming a ladder leg [Fig.~\ref{sites}(b)] (the {\it leg sites} hereafter). (III) In the third class of site [Fig.~\ref{sites}(c)] the muon sits within a CuBr$_4$ tetrahedron (the {\it tetrahedron sites}, hereafter). 
These three stopping sites: rung, ladder and tetrahedron,  have many features in common.  Most notably, in all cases the muon sits between two Br atoms.  This is true even for the case in which the muon lies inside the CuBr$_4$, where the muon sits along an edge of the tetrahedron. We conclude that the Br-$\mu^{+}$-Br-like state is highly probable for a muon stopping in this material, with  the muon consistently found to sit slightly closer to one of the Br atoms than the other. 

Most importantly for our discussion, each of these sites causes significant structural, electronic and magnetic distortion to the local environment (see Appendix).
In Fig.~\ref{sites} we show the most significant atomic displacements.  The lighter spheres indicate the initial positions of the atoms whereas the darker spheres represent the relaxed positions of the atoms.  The displacements of all other atoms (not shown) are less than 0.2~\AA.
In all cases we observe a significant displacement of the nearest Cu atom, which is accompanied by a stretching of the nearest Cu--Br bond.  In the leg and rung sites it is the Cu atom within the tetrahedron containing the Br atom that the muon sits closest to that experiences a significant displacement.   
The perturbation on the local magnetism caused by the muon (and described in the Appendix) is also significant, involving a reduction in the local Cu moments and changes in their local spin polarization over sizeable distances.

\subsection{Mapping the features in the TF spectra}

We can consider the three muon sites identified above in reconciling the features seen in the TF spectra, where we have four types of muon state: two experiencing local ranges of magnetic fields close to the magnitude of the applied magnetic field (i.e.\ they see  a small average internal magnetic field contribution from the sample) and one either side of these (which experience a larger internal field contribution). Below we discuss how each of these muon states predicted by DFT might contribute to the observed features in the spectra. We argue that the features observed on either side of the central peak are likely to be due to a sizeable perturbation to the system caused by the muon-induced distortion to the system with associated local staggered moments, and identify the rung and leg sites as those most likely to be responsible for these features, owing to the disruption to exchange couplings that they are predicted to cause in their vicinity. (Note that the structure of the material suggests that there are two exchange pathways along the rung and one along the leg, with the two rung paths related by an inversion center. If the muon is close to one Br, we then expect to find two muon sites coupled to the rungs and two sites coupled to the legs.)

The observed central peaks are of rather different character to the peaks we observe on either side: one central contribution is relatively narrow and one rather broad. The narrow central peak varies relatively little over much of the field range and would appear to result from muon sites not well coupled to the material itself.
 Sites contributing to this latter peak might include muons stopping outside the sample in the sample holder, or from muons in sites in the sample where they do not experience a significant static internal local magnetic field.  
 We do not discuss these sites further as they provide little information about the system.  The broader peak is better coupled to the magnetic behaviour, showing a maximum in its width on passing through $B_{\mathrm{c}}$ which suggests a dynamical origin.  Sites in which a muon sits close to the (C$_5$H$_{12}$N)$^{+}$ ions might reasonably produce these central peaks as a result of the muon sitting further away from the Cu ions and thereby experiencing a relatively small internal local field.  However, such sites were not found in our calculations, even with the muon being initialised close to the piperidinium rings, with the exception of one site which was found to be around 2~eV higher in energy than the low energy muon sites.  We therefore conclude that such a site is unlikely to be realised.  This is to be contrasted with calculations performed on the strong-leg spin ladder system (C$_5$H$_9$NH$_3$)CuBr$_4$ \cite{willet} which demonstrate sites where the muon was able to form a C-$\mu^+$ bond, breaking the ring.   Calculations performed on an isolated piperidinium ion showed a similar ring-breaking and bond formation.  This implies that the formation of the $\mu^+$-Br bond is so energetically favourable in the current case that muon sites not involving this coordination are unlikely to occur.
In the absence of sites near the carbon atoms, we tentatively attribute the broad central peak to muons stopping in tetrahedron Br-$\mu^{+}$-Br sites. 
Whereas the perturbation of the rung and leg sites is identified as being necessary for the associated spectral features we observe, muons in the tetrahedron sites do not sit directly along the magnetic exchange pathways and thus might be expected to perturb the system much less strongly than the rung and leg sites.

The two features appearing on either side of the central lines result from muon sites that couple to the physics of the spin ladders,  as the features correlate strongly  with the known phases of the material. 
In the QD region we observe a feature shifted to the high-field side of the central peak, while in the LL/LRO regions we see a feature on the low field side. 
Note that the discontinuity in the field shift on passing through $B_{\mathrm{c}}$, the difference in slope of their field dependence [Fig.~\ref{fig:fitting-undoped}(a)] along with the appearance of both features simultaneously in the QC region would strongly suggest that they do not arise from a single feature in the spectra that moves continuously as a function of applied field, indicating that they each require separate explanations. 
If we consider a muon site in the crystal and the region of the material around it, 
 the total local magnetic field experienced by the muon will be the vector sum of the externally applied field $B_{0}$, internal field from the surrounding pocket of material $B_{\mathrm{int}}$, the Lorentz field $B_{\mathrm{L}}$ from more distant moments and the demagnetizing field $B_{\mathrm{demag}}$. 
Approximating the sample as a flat plate, we would expect that $B_{\mathrm{demag}} = - \mu_{0}M$, where $M$ is the magnetization, and that the Lorentz field will be given by $B_{\mathrm{L}}=\mu_{0}M/3$. 
The field at the muon site is then $B = B_{0} -2\mu_{0}M/3 + B_{\mathrm{int}}$. 

The unperturbed QD region, formed from interacting antiferromagnetic dimers, should give no internal magnetic field at the muon site arising from the Cu$^{2+}$ spins and should have magnetization $M=0$ and so we would therefore expect $B=B_{0}$. The unique signal we observe on the high-field side of the central peak in this phase must therefore result from a small additional contribution to the magnetization or internal field. 
One possible source of a small magnetization identified in the BEC compound DTN\cite{sjb} was that a
small misalignment of the crystal with respect to the applied magnetic field can lead to a small non-zero magnetization $M$, however no such effect has been reported for our material.
In order to observe a peak at positive $B-B_{0}$, we require a positive contribution to $B_{\mathrm{int}}$ in a region of the phase diagram where there should be no electronic spins in the unperturbed state. The main candidate for producing such an effect is a muon stopping site such as the rung or ladder state  that distorts its local environment to the extent that a hyperfine or dipolar contribution from the local Cu$^{2+}$ moments becomes resolvable. This might, for example, involve the rung site disrupting the  local exchange pathway between the two Cu$^{2+}$ ions in a dimer, or a more general distortion leading to a hyperfine field at the muon site. 
One rather extreme example of a muon induced distortion, still consistent with the DFT result, would be the breaking of a local dimer by the muon which would result in an unpaired spin near the muon site aligned preferentially along the applied field $B_{0}$. 
Such a distortion was also suggested to be the case in $\mu^{+}$SR measurements on the double chain compound KCuCl$_{3}$ \cite{andreica}. 
There a TF signal comprising seven frequencies was observed at low temperature with a temperature dependence suggestive of unpaired Cu$^{2+}$ spins. However no comparable spectral structure or temperature dependence is observed here. 

\subsection{Behaviour in LL/LRO phases}

We now turn to the features on the low field side of the central peak that are most evident in the LL and LRO phases. 
In the LRO phase in the spin ladders \cite{thielemann} at $8.6$~T, the spins are aligned
perpendicular to the $a$-axis and antiparallel within the ladder,
but parallel on ladders of the same type [propagation vector
$k=(0.5, 0, 0)$]. The ordered moment is 0.41~$\mu_{\mathrm{B}}$ per copper
ion.
Increasing  the applied field in the region $B_{0}>B_{\mathrm{c}}$ cants the AF order parallel to $B_{0}$. 
Thus the magnetization rises
rapidly and via the negative contribution
from the sum of the Lorentz and demagnetization
fields, should be expected to result in the appearance of a new peak
at low field which moves to still lower frequency as the field is further increased, as we observe. 
Thus, the low-field peak
which is seen at fields exceeding $B_{\mathrm{c}}$ has a separation
from the central peak which increases proportionally
not to $B_{0}$ but to $B_{0}-B_{\mathrm{c}}$ and thus tracks the chemical
potential of the bosonic excitations. 

However, just as in the QD region, it is unlikely that the muon probe coupling passively to the field of the local magnetic moments provides
 the entire explanation here.  This is because the low field feature is seen not just in LRO phase but also in the LL phase,  where there should not be magnetic order. 
In the LL phase there will be some magnetization owing to correlations that locally resemble the spin configuration in the ordered phase, leading to a larger contribution $-2\mu_{0}M/3$ from the Lorentz and demagnetizing fields. 
However in this phase the local magnetic field will be rapidly fluctuating on the muon time scale. 
Moreover, the distortions close to the muon described above will still be active and might be expected to produce a distortion in the local magnetism of the LL electronic state, as has been previously suggested theoretically \cite{eggert,eggert2}. 
In fact, there is experimental evidence for such states in the low temperature behaviour of
 the 1DQHAF dichlorobis (pyridine) copper (II) (CuCl$_{2}\cdot$ 2NC$_{5}$H$_{5}$ or CPC) \cite{chakhalian}. 
In this case, shifts in the TF $\mu^{+}$SR spectra were attributed to muon-induced perturbations to the spin chain. 
The physics here involves the muon moment causing a significant perturbation to the local exchange links, leading to a 
local susceptibility which differs markedly from that of the rest of the chain \cite{eggert,eggert2}. In the case of CPC, where a powder sample was measured, peaks were resolved on both sides of a central peak, attributed to site dependence of the sign of the hyperfine field, with the fact that the sample was a powder causing a dipolar broadening only, and no shift in the peak position.

The occurrence of both satellite features in the QC region provides evidence that the low field feature does not simply track the macroscopic magnetization, since the appearance of the low-field feature below $B_{\mathrm{c}}$ suggests that the muons are sensitive to LL correlations and that these start to form on the muon response scale in the $6 \leq B \leq 7$~T region. 

At this point we can turn to the calculations of local field at the three sites outlined in the appendix. These indicate that a straightforward assignment of the satellite features can be made, the high field feature being from rung sites in an environment with quasistatic local canted AF order and the low field satellite from leg sites within the same quasistatic local canted AF environment.  
The presence or absence of each of these features in the different phases reflects an interplay between the muon-induced perturbation at each site and the underlying state of the system.


As noted above, impurity-induced local AF order is a known property of chain systems, thus its presence in response to the more strongly perturbing muon sites across a large part of the phase diagram is not entirely surprising. A key question is why such an effect disappears for the rung site in the LL/LRO phase as well as for the leg site in the QD phase. 
One possibility (examined further below in section \ref{tf2}) is that the presence of the muon in the rung sites site leads to a distortion that reduces the size of the local spin gap. This would explain the appearance of the low field feature at fields below $B_{c}$ and also to the occurrence of a feature in the QD region.
Another possible explanation for the disappearance of the static features is that dynamical interaction between the muon-induced moments and the QD and LL/LRO phases leads to fast spin fluctuation and removes the static component of the field at the respective muon sites in these phases. The precise mechanism for this behaviour is unclear at present and it is clearly an area that would benefit from further theoretical investigation.

In conclusion, it is difficult to reconcile the features we see in the spectra without invoking a significant distortion to the local electronic structure caused by the implanted muon. The two stopping sites lying along the exchange pathways, both found via our DFT calculations, provide such distortions. However even in the presence of this local distortion, we obtain here the striking and important result that the nature of the resulting stopping state allows the muon to probe the global, underlying physics of the spin ladder via its response to the perturbing muon. 


\section{TF $\mu^{+}$SR on  (H\lowercase{pip})$_{2}$C\lowercase{u}B\lowercase{r}$_{4(1-x)}$C\lowercase{l}$_{4x}$}\label{tf2}

Using the conclusion from the previous section that features in the TF $\mu^{+}$SR spectra may be correlated with the regimes of behaviour in spin ladder systems, the natural extension is to investigate the possibility of new phases in disordered spin ladder systems such as  (Hpip)$_{2}$CuBr$_{4(1-x)}$Cl$_{4x}$. 
In
the itinerant boson picture, the controlled introduction of disorder that we investigate here has a dramatic impact on the single-particle eigenstates for the bosons, leading to the possibility of new phases such as the Bose glass phase in the dilute doping limit. The disorder introduced by the randomized exchange
interaction strengths  acts to locally reduce the singlet-triplet gap, so that for fields below $B_{\mathrm{c}}$ local condensation of triplets with finite susceptibility may occur. 
Specifically, in the case of this material, the substitution of Br for Cl is expected to influence both the rung and leg exchange pathways. If a rung bond is diluted, for example,  two localised moments are created that still couple antiferromagnetically,
but with a smaller exchange, which could lead to a local closing of the energy gap \cite{ryll_thesis}.

\begin{figure}
\begin{center}
\epsfig{file=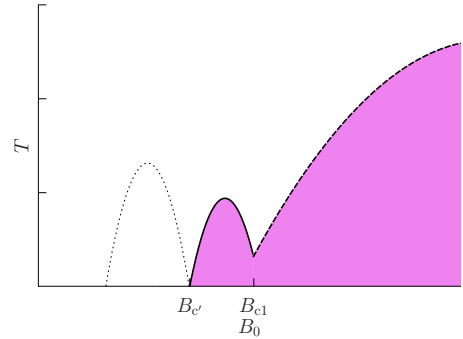, width=7cm}
\caption{Suggested phase diagram for disordered (Hpip)$_{2}$CuBr$_{4(1-x)}$Cl$_{4x}$ on the basis of thermodynamic measurements of the $x=0.1$ concentration \cite{ryll_thesis} and our $\mu^{+}$SR measurements of the $x=0.05$ concentration (see main text). An additional dome appears between $B_{c\prime}$ and $B_{c1}$ and a coexisting magnetic phase with distinct relaxation parameters is indicated (dotted line).
\label{fig:phasediagram2}}
\end{center}
\end{figure}

Specific heat and magnetocaloric effect (MCE) measurements on members of the doped system with $x=0.1, 0.5$ and $0.9$ were reported in Ref.~\onlinecite{ryll_thesis}.
Our measurements are made on a sample with $x=0.05$ and so we might expect a phase diagram that is similar to that observed in the $x=0.1$ case. 
 For the case of the thermodynamic measurements on $x=0.1$, the typical dome-like structure of the Luttinger liquid regime is observed in roughly the same place,  in addition to a new feature in the phase diagram in the form of a smaller dome centred around 6.5~T with an onset at a lower critical field $B_{\mathrm{c'}} = 6.1(3)$~T [shown schematically in Fig.~\ref{fig:phasediagram2}]. MCE measurements suggest a sharp cusp separating the two domes at $B_{\mathrm{c1}}= 7.1(2)$~T. In addition, heat capacity measurements suggested  that around $4 \lesssim B_{0} \lesssim 6$~T, there is a region depicted using dotted lines in Fig.~\ref{fig:phasediagram2}, whose temperature relaxes differently with time to that in the rest of the phase diagram (it requires an additional exponential relaxation for fitting the heat capacity response).  This effect was suggested to be indicative of a coexisting magnetic system whose interaction with the phonon system acts on a different time scale. 
As pointed out in Ref.~\onlinecite{ryll_thesis}, the resulting phase diagram bears a resemblance to that predicted for the case of 3D dimer system on the basis of quantum Monte Carlo simulations \cite{nohadani}. Here two dome-like regions are predicted: one at higher field involving the BEC of all spins and a lower-field dome reflecting the microcondensation of triplons on disordered bonds. A Bose glass phase is predicted at fields below each of the domes. In our case, the coalescence of the domes only leaves the possibility of glassy  in the region at fields below  a critical field $B_{\mathrm{c'}}$.

\begin{figure}
\begin{center}
\epsfig{file=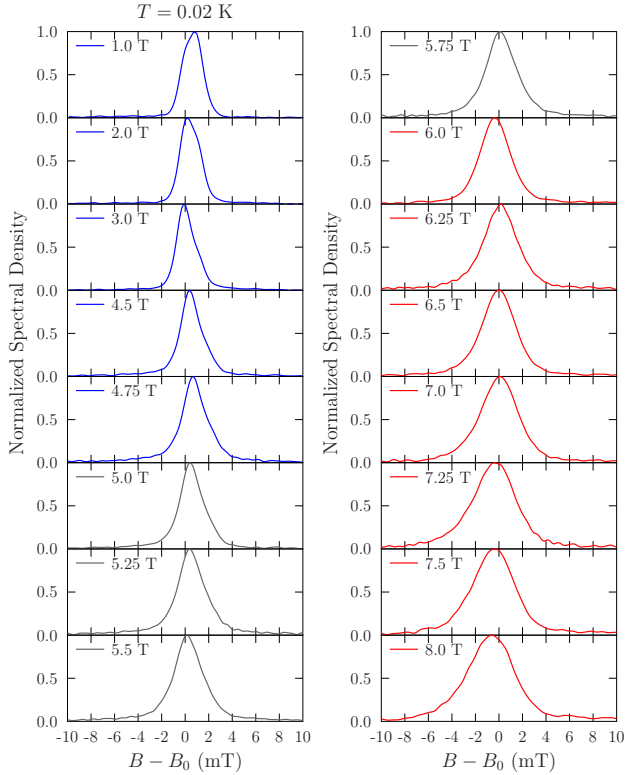, width=\columnwidth}
\caption{Fourier transform TF $\mu^{+}$SR spectra measured for (Hpip)$_{2}$CuBr$_{4(1-x)}$Cl$_{4x}$ ($x=0.05$) at  $T=0.02$~K.  \label{fig:spectra2}}
\end{center}
\end{figure}
\begin{figure}
\begin{center}
\epsfig{file=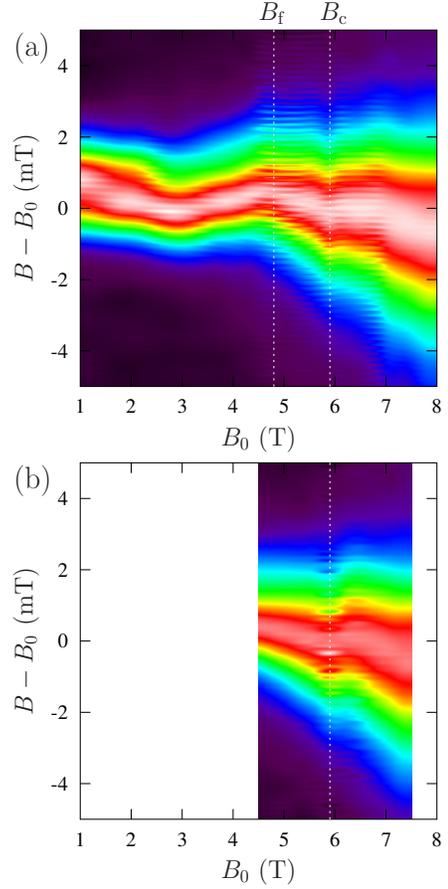, width=6.5cm}
\caption{A color map of the the Fourier amplitude as a function of applied field $B_{0}$ for TF measurements on (Hpip)$_{2}$CuBr$_{4(1-x)}$Cl$_{4x}$ with $x=0.05$.
Data are shown at (a) $T=0.02$~K and (b) $T=0.7$~K. 
White lines are guides to the eye showing probable transitions (see text). \label{fig:contours2}}
\end{center}
\end{figure}

Example Fourier transform TF $\mu^+$SR spectra are shown for  (Hpip)$_{2}$CuBr$_{4(1-x)}$Cl$_{4x}$ with $x=0.05$ in Fig.~\ref{fig:spectra2}, measured at $T=0.02$~K. In contrast to the $x=0$ case described in Section~\ref{tf}, the central line is rather broad with no additional peaks  resolvable as a function of applied field. However, the shape and position of the line can be seen to change both in Fig.~\ref{fig:spectra2} and more clearly in the color map plots shown in 
Fig.~\ref{fig:contours2}. Particularly notable in the color maps are marked changes seen at $B_{0}=5.9$~T (at both $T=0.02$ and 0.7~K) and at $B_{0}=4.8$~T (seen only in the $T=0.02$~K scan). 

\begin{figure}
\begin{center}
\epsfig{file=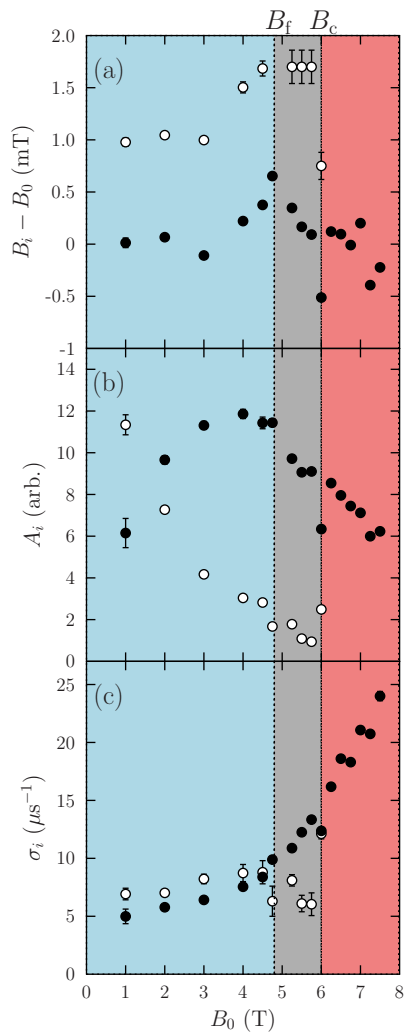, width=6.cm}
\caption{Results of fitting  Eq.~\ref{fit2} to the data measured for the $x=0.05$ compound at $T=0.02$~K. (a) Peak position $B_{i}$ relative to the applied field $B_{0}$. (b) Amplitudes $A_{i}$. (c) Peak widths $\sigma_{i}$. \label{fig:fitting-doped}}
\end{center}
\end{figure}

The broad line observed in the spectra 
made the fitting of several Gaussian peaks impossible 
for this material. Instead, it was found that the features were best fitted by a sum of two broad Lorentzian peaks using Eq.~\ref{fit2}, with $N=2$. 
 (The success of the Lorentzian lineshapes in fitting the spectra probably reflect  the sum of several  components broadened significantly compared to the $x=0$ case.)
The results of fitting the spectra to this function are shown for measurements at $T=0.02$~K in Fig.~\ref{fig:fitting-doped}. As in the color maps, marked changes are observable in the fitted parameters at $B_{0}=4.8$~T and 5.9~T.  

As in the $x=0$ case we have evidence for a high-field feature whose amplitude decreases continuously [Fig.~\ref{fig:fitting-doped}(b)] and  becomes very small above $B_{0}=5.9$~T, with a shift in field that peaks in the region $4.8 < B_{0}<5.9$~T. This would seem to correspond to the high field peak seen in the $x=0$ material in the QD phase, although it is not resolvable as a separate line, owing to the width of the central line. 
The central feature is seen to increase in amplitude in the QD regime [Fig.~\ref{fig:fitting-doped}(b)], and then decrease continuously from around $B_{0}=4.8$~T, while its width increases monotonically with increasing field [Fig.~\ref{fig:fitting-doped}(c)]. 
The centre of both features shifts with field [Fig.~\ref{fig:fitting-doped}(a)] showing quite large changes in the shift direction at $B_{0}=4.8$~T and $B_{0}=5.9$~T. 

The  disappearance with increasing $B_{0}$ of the high field component in the $x=0$ concentration and the present material allows us to suggest $B_{c'}=5.9$~T for the $x=0.05$ material, similar to that observed in $x=0.1$.  
Given the  change in  around $B_{\mathrm{f}}=4.8$~T, we also have the possibility of a crossover into a new region of behaviour for the field range $4.8 < B_{0}<5.9$~T at $T=0.02$~K. (Note that at these low temperatures, we would not expect a wide region of QC-like behaviour.) This is the part of the phase diagram where glass-like behaviour is expected.  In this range of applied field, the high field component persists but at low amplitude, while the central line shifts back towards $B_{0}$. This region is therefore characterized in our $\mu^{+}$SR measurements by behaviour intermediate between that observed in the QD region and the  LL region. This is suggestive of a division in the muon sites between environments showing each of these behaviours. 
This scenario is in good agreement with the observation of the heat capacity taking on a two-exponential form in the $4\lesssim B_{0}\lesssim 5.8$~T region in the $x=0.1$ material, which is characteristic of two coexisting magnetic states \cite{ryll_thesis}. 
Moreover, one of the predictions of Bose glass physics is a scenario where locally the gap in the excitation spectrum closes at disordered sites, producing the same environment as seen in the LL phase \cite{zheludev}. We therefore suggest that something similar may be happening in the $4.8 < B_{0}<5.9$~T region. It is notable that the  features that we see at $T=0.02$~K are absent in the data measured at $T=0.7$~K, leading us to suggest that the situation may look like the phase diagram shown in Fig.~\ref{fig:phasediagram2} with the peak in the glassy dome occurring below 0.7~K, in agreement with that inferred from specific heat \cite{ryll_thesis}.

\section{Longitudinal field measurements}\label{lf}

Using LF $\mu^{+}$SR allows us to investigate the excitations of the spin ladder system on the muon ($\mu$s) timescale. 
In such measurements, the spin relaxation is determined by the magnetic field correlations perpendicular to the applied magnetic field. 
We make LF measurements on (Hpip)$_{2}$CuCl$_{4}$ which has $B_c$ lying within the field range of the HiFi spectrometer at ISIS. 
On the basis of our DFT studies of (Hpip)$_{2}$CuBr$_{4}$ we expect the (Hpip)$_{2}$CuCl$_{4}$ system will also have a tetrahedron site that will dominate the LF relaxation.
LL theory provides a complete quantum mechanical treatment of the low temperature behaviour for $B_{0}>B_{\mathrm{c}}$ built upon only two interaction-dependent parameters: $u$ and $K$. 
Dynamical spin excitations within the LL are best viewed as being due to interacting spinless fermions, where the LL parameter $K$ is a measure of the sign and strength of the interactions. 
NMR investigations  \cite{klanjsek}  of the spin-lattice relaxation rate $1/T_{1}$ 
led to the prediction
$1/T_1 (T) \propto T^{(1/2K) -1}$, which allows the determination of $K$ from the power law exponent $\alpha = (1/2K)-1$, and therefore provides a quantitative test of LL theory that we can access with our measurements. 

In order to test whether longitudinal field (LF) $\mu^{+}$SR could be used to probe the dynamics of the spin ladder system, measurements were previously performed \cite{moeller2} on the strong-leg spin ladder material (C$_7$H$_{10}$N)$_2$CuBr$_4$ (known as DIMPY) using the HiFi instrument at the ISIS facility.
DIMPY has $B_{\mathrm{c}}\approx 3.0$~T and, at applied fields above $B_{\mathrm{c}}$,  orders below $T_{\mathrm{c}}\approx0.34$~K.
As a result of lying in the strong-leg regime, DIMPY has a Luttinger parameter $K>1$, which corresponds to attractive interactions between fermionic excitations. This leads to a prediction that the power law exponent $\alpha<-0.5$.
It is notable that this scaling applies in the LL regime only, so we require not only that $B_{0}>B_{\mathrm{c}}$ but also that $T_{\mathrm{c}}\ll T\ll J$ and this was borne out in the NMR results\cite{jeong}, where the predicted behaviour 
was found in the realm of applicability of the model. 
\begin{figure}
\begin{center}
\epsfig{file=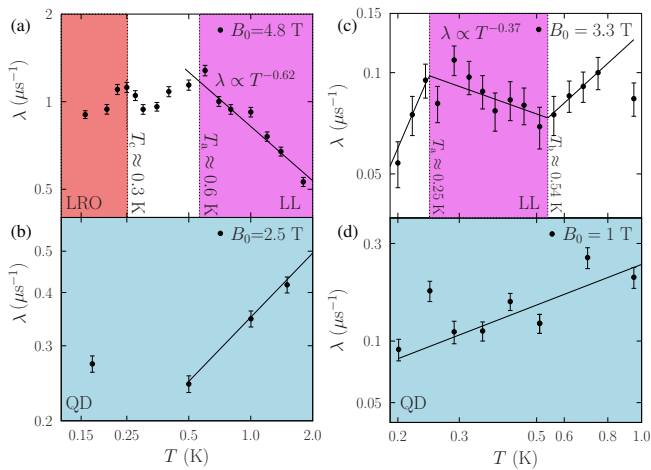, width=\columnwidth}
\caption{Temperature dependence of the LF $\mu^{+}$SR relaxation rate at fixed applied longitudinal magnetic field $B_{0}$. (a,b) The strong leg spin ladder DIMPY. (c,d) The strong rung spin ladder (Hpip)$_{2}$CuCl$_{4}$. \label{fig:dynamics}}
\end{center}
\end{figure}

The measured spectra for both DIMPY and (Hpip)$_{2}$CuCl$_{4}$ were fitted to the relaxation function
\begin{equation}
A(t) = A_{\mathrm{rel}}{\rm e}^{-\lambda t} + A_{\mathrm{bg}},
\end{equation}
where $A_{\mathrm{rel}}$ is the fixed, relaxing asymmetry and $A_{\mathrm{bg}}$ is a field-dependent nonrelaxing component. 
The observation of a single relaxation rate suggests that muons from one class of site contribute most strongly to the observed dynamics. The fitted relaxation rate might  be expected to principally reflect the sites that give the broad central component in the TF data, since these makes the largest contribution of the three sites that are sensitive to the intrinsic magnetism of the system. The relaxing amplitude for  (Hpip)$_{2}$CuCl$_{4}$ was found to be relatively small at a few \%, reflecting the constraints of the instrument and sample geometry and the relatively small fraction of muons with strong enough coupling not to have their relaxation quenched by the LF. The temperature dependence of the slow relaxation rate at fixed field is the parameter of interest here and we did not expect or detect any significant temperature dependence of $A_{\mathrm{rel}}$, so it was held fixed in the fitting. 

At the magnetic fields of interest we expect any nuclear contribution to the muon relaxation to be fully quenched and therefore we probe only the electronic spin dynamics. \cite{moeller2}
 The LF relaxation rate $\lambda$ is then equivalent to $1/T_{1}$ measured in magnetic resonance. 
Fig.~\ref{fig:dynamics}(a,b) shows the $\mu^{+}$SR relaxation rate $\lambda (T, B_{0})$  found previously\cite{moeller2} for measurements on DIMPY as a function of temperature for fixed values of applied field $B_{0}$. 
For the QD regime, realized for $B_{0}<B_{\mathrm{c}}$ [Fig.~\ref{fig:dynamics}(b)],
the muon relaxation rate $\lambda$ is seen to increase with  temperatures above 0.5~K. For the LL regime, realized for  $B_{0}>B_{\mathrm{c}}$, [Fig.~\ref{fig:dynamics}(a)] a  local maximum is seen in the relaxation close to $T_{\mathrm{c}}$, followed by an increase in $\lambda$  with increasing temperature,  with a  power law decrease found at temperatures significantly above $T_{\mathrm{c}}$ (but well below $J$). This was in good agreement with the NMR $1/T_{1}$ results, where the scaling only begins to be seen above $T_{\mathrm{a}} \approx 2T_{\mathrm{c}}$ which suggests an onset temperature of $ T_{\mathrm{a}} \approx 0.6$~K for muon measurements made at $B_{0}=4.8$~T, in good agreement with that shown in Fig.~\ref{fig:dynamics}(a).
The exponent $\alpha$,  determined by fitting $\lambda \propto T^{\alpha}$,   was found to be $\alpha = -0.62(5)$ at $B_{0} =4.8$~T which is comparable to the NMR values, which showed that  $\alpha$ decreases rapidly in the $3.5 < B < 5$~T range, settling at $\alpha=-0.75$ for fields above 9~T (inaccessible in these measurements, which are limited to $B_{0}<5$~T). 

In contrast to DIMPY, (Hpip)$_2$CuCl$_4$, which has $B_{\mathrm{c}}\approx 1.7$~T, lies within the strong-rung regime and  so is expected to host dynamics that can be represented in terms of repulsive fermionic excitations with $K<1$.  Correspondingly, we expect a value of $=-0.5 \leq \alpha \leq 0$. Results of our LF measurements are shown in Fig.~\ref{fig:dynamics} (c,d). In the QD regime at $B_{0}=1$~T [Fig.~\ref{fig:dynamics}(d)], the relaxation rate again tends to increase with increasing temperature. 
The behaviour is more complex at $B_{0}=3.3$~T, but is found to be consistent with the data measured for DIMPY, albeit in a slightly lower temperature regime. 
By analogy with the Br-containing material, we would expect 
the top of the LL dome coherence energy in (Hpip)$_2$CuCl$_4$ to be around $J\approx 0.6$~K, with a prediction that the scaling breaks down as we leave the realm of applicability of the model. 
If we also assume a $T_{\mathrm{c}} \lesssim 0.1$~K, then we should expect power law  for $0.2 \lesssim T \lesssim 0.6$~K. Finally, by analogy with DIMPY, we might expect that $\lambda$ increases with increasing temperature at temperatures below the scaling regime. 
Such a low temperature increase in $\lambda$ is indeed observed, followed by a regime of power law behaviour, followed by a further increase in $\lambda$ for $T \gtrsim 0.6$~K.
Fitting the data in the range $0.25< T < 0.6$~K suggests $\alpha=-0.35\pm 0.15$, putting the material in the repulsive regime as expected. 
Motivated by the fact that the data suggests three regimes of behaviour, Fig.~\ref{fig:dynamics}(c) shows the result of fitting a model that allows three different power laws with sharp crossover temperatures, whose values are also parameters in the fit. This results in crossover temperatures of $T_{\mathrm{a}}=0.25$~K and $T_{\mathrm{b}}=0.54$~K, with a power law in the LL regime of $\alpha=-0.37 \pm 0.10$. 
We may conclude that, as in the case of DIMPY, the scaling behaviour is quite dependent on the precise position within the $B$-$T$ phase diagram. Although this gives us reason to be cautious about the precise value of the power law,  the results
are consistent with the range $-0.5 \leq \alpha \leq 0$,  expected for a strong rung ladder. 

\section{Conclusions}
We have presented a broad survey of measurements made on spin ladder systems using implanted muons. 
The transverse field technique has been shown to be sensitive to the crossover between quantum disordered and Luttinger Liquid regimes in (Hpip)$_{2}$CuBr$_{4}$. 
The muon probes the magnetism of these systems by realizing a local magnetic perturbation that results from the distortion to the local structure caused by its electrostatic charge. Despite this modified local character, the muon continues to be a useful probe of the global properties of the system, 
enabling the phase diagram to be mapped out across a wide range of field and temperature.

This allows us to probe the phase diagram of the partially disordered system  (Hpip)$_{2}$CuBr$_{4(1-x)}$Cl$_{4x}$ ($x=0.05$), where we identify a new regime of behaviour in the region of applied fields $4.8 \leq B_{0}\leq 5.9$~T which, locally, is intermediate between the QD and LL regions, broadly consistent with predictions of Bose glass descriptions of the physics in related systems. In addition,  longitudinal field measurements enable us to probe the spin dynamics of the ladders and distinguish the dynamics of strong-rung and strong-leg materials via characteristic Luttinger liquid parameters. 

These results demonstrate the use of muons as a local probe in higher magnetic fields than are typically employed in standard $\mu$SR studies and provide us with an insight into the evolution of behaviour between systems based on interacting spin dimers such as the putative spin liquid states. It is possible that the muon causes significant perturbations to the global magnetic state via the local magnetism  in such spin liquid systems (and many others). However, the present results show that these are not necessarily detrimental to the use of the muon as a probe of the broader intrinsic properties, but instead provide a means of probing them via the local magnetism. This is especially important in quantum disordered phases, where in the absence of such a local perturbative probe effect we would expect no magnetic response at all. It is hoped that in future the detailed analysis of muon stopping states in an increasingly diverse range of materials will provide further insight into the ways in which the muon probes its host material, thus allowing a richer level of detail on the properties of the material to be revealed. 

\section{Acknowledgements}
Part of the work was carried out at the Swiss Muon Source, Paul Scherrer Institute, Switzerland and at the STFC ISIS Facility, Rutherford Appleton Laboratory, UK and we are grateful for the provision of beamtime. 
We thank Thomas Hicken for useful discussions. The work was supported by EPSRC grants EP/N024028/1, EP/N023803/1 and EP/N024486/1 and by NSF grant DMR-1703003. Fan Xiao thanks the John Templeton Foundation for support via a fellowship. Ben Huddart thanks STFC and Rob Williams thanks EPSRC for support via studentships.  Data presented here will be made available via the Durham University data archive.

\appendix*

\section{Density functional theory calculations}

The proposed muon stopping sites were obtained from structural relaxations of the system plus an implanted muon using DFT. The muon is modelled as an ultrasoft hydrogen pseudopotential.  We use a plane wave cut off energy of 1000~eV resulting in energies that converge to a precision of $\sim 10$~meV per cell and perform Brillouin zone integration at the $\Gamma$ point. We consider the case of $\mu^+$ by using a charged cell and use a neutral cell to study the  of muonium, formed when the muon attracts an electron as it moves through the crystal.

\begin{figure}[ht]
	\includegraphics[width=\columnwidth, trim={0cm 0cm 0cm 0cm},clip]{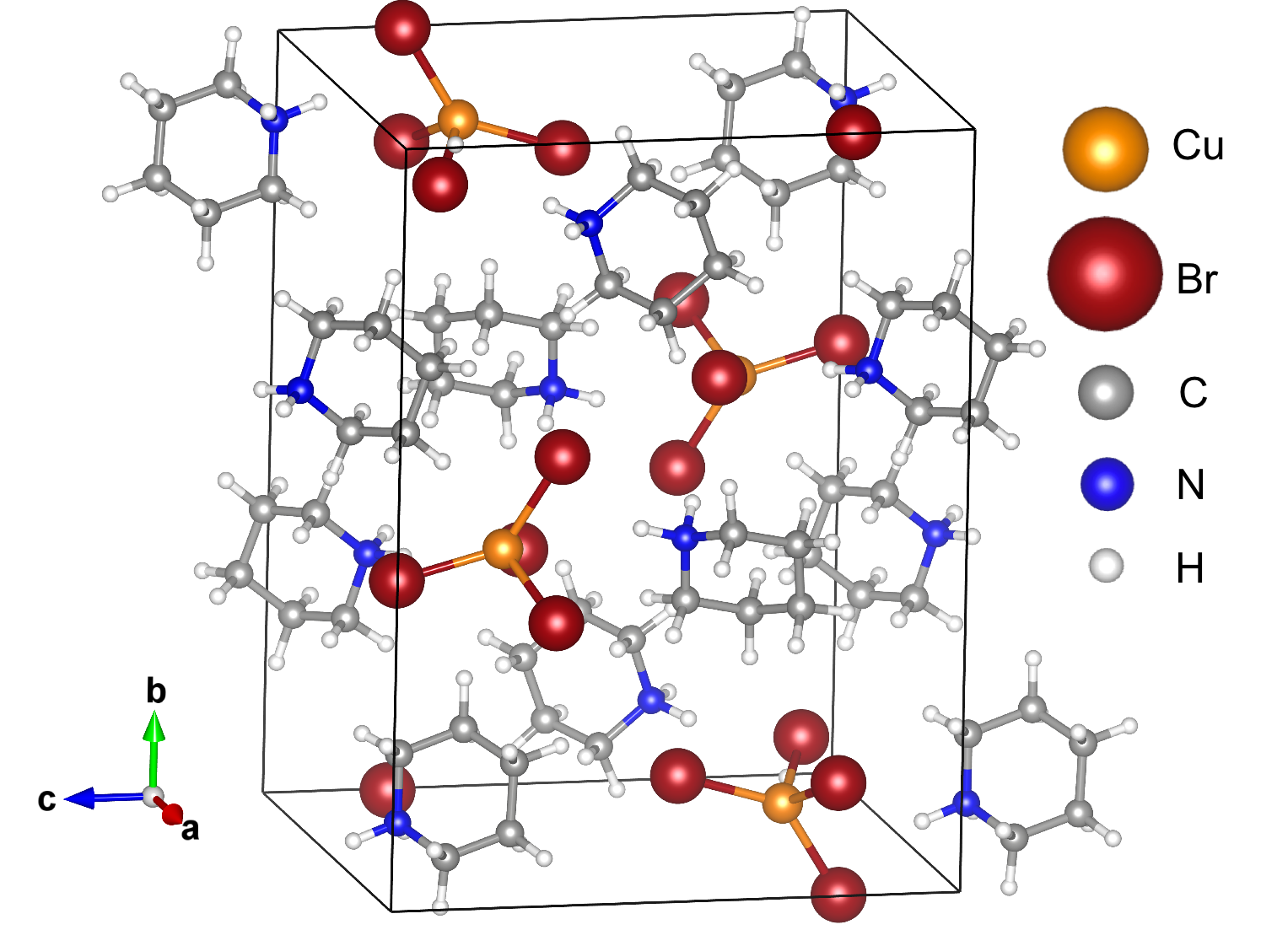}
	\caption{The unit cell of $(\textrm{Hpip})_2 \textrm{CuBr}_4$, comprising (CuBr$^{2-})_4$ tetrahedra and (C$_5$H$_{12}$N)$^{+}$ counterions.}
	\label{Hpip_all}
\end{figure}

$(\textrm{Hpip})_2 \textrm{CuBr}_4$ crystallizes in the monoclinic space group P2$_1$/c space group \cite{willet}, with $a=8.487(2)$~\AA, $b=17.225(3)$~\AA~$c=12.380(2)$~\AA \ and $\beta=99.29(2)^{\circ}$.  The unit cell comprises flattened (CuBr$^{2-})_{4}$ tetrahedra and (C$_5$H$_{12}$N)$^{+}$ counterions and is shown in Fig.~\ref{Hpip_all}.  Structural relaxations were performed using a $2 \times 1 \times 1$ super cell.  Doubling the simulation cell along the shortest dimension reduces the spurious self-interaction of the muon which results from the use of periodic boundary conditions.  Furthermore, this results in a simulation cell that is equivalent to the magnetic unit cell of this system, thereby allowing us to better assess the effect of the implanted muon on the magnetic structure.

The muon was placed in 52 different initial positions (for both charged and neutral cells) forming an equally spaced three-dimensional grid spanning distinct positions within the conventional unit cell.  Positions where the muon would be $<1$~\AA\ away from another atom were discarded as starting points.  The structure plus implanted muon was then allowed to relax until the forces on the atoms were all $< 5 \times 10^{2}$~eV/\AA\ and the total energy and atomic positions converged to $2 \times 10^{-5}$ eV and $1 \times 10^{-3}$~\AA\ respectively.  We also relaxed the structure without the muon in the same manner, such that any atomic displacements result from the presence of the muon, rather than differences between the experimental structures and that obtained from DFT.

\begin{figure}[ht]
	\includegraphics[width=\columnwidth ]{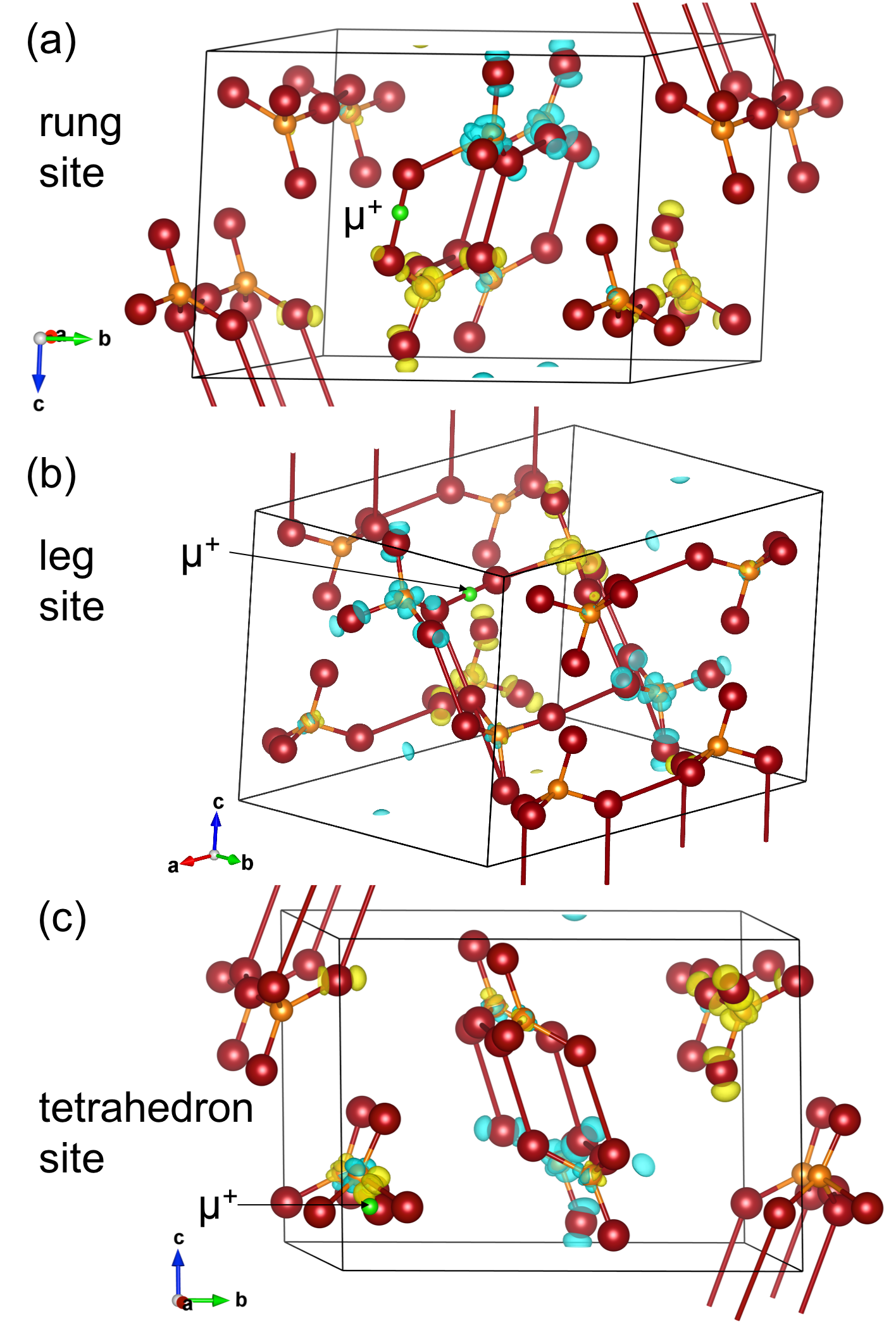}
	\caption{Muon stopping sites within the spin ladder structure.  Muons were found to stop (a) along ladder rungs (b) along ladder legs and (c) inside the CuBr$_4$ tetrahedra.  Yellow and cyan isosurfaces indicate increases and decreases in spin density respectively.}
	\label{sites2}
\end{figure}

We first consider the case of diamagnetic $\mu^+$. Relaxation of the conventional cell plus the implanted muon yields, as might be expected for a molecular material, many candidate muon sites that are close in energy. 
We find that the muon stops close to the electronegative Br ions in the CuBr$_4$ tetrahedra as detailed in the main text. Rung sites [Fig.~\ref{sites2}(a)] have the lowest energies lying within a narrow 30 meV range.   Leg sites [Fig.~\ref{sites2}(b)] were found to be on average 76 meV  higher in energy than the rung sites. Many tetrahedral sites [Fig.~\ref{sites2}(b)] were identified, having energies 43--204~meV higher than the rung sites.  
 
Since changes in spin density induced by the implanted muon are likely to have a significant impact on the magnetic properties measured in a $\mu^{+}$SR experiment, we have compared the calculated spin density of the structure with and without the muon.  It is found experimentally that the magnetic exchange coupling is antiferromagnetic along both the legs and rungs and we therefore expect the sign of the spin density to alternate between adjacent Cu ions, as found by neutron diffraction\cite{thielemann}. 
We note, however, that this is not the magnetic ground state found in our calculations, even for the unperturbed structure, where we find a complicated magnetic ground state involving 
spins of equal magnitude on each Cu site, with  mixed ferromagnetic and antiferromagnetic coupling within the ladders.  The failure of DFT to correctly capture the ordered magnetic ground state of this system is perhaps unsurprising, given that the system is found to order only upon the application of a relatively large external magnetic field.
Though this fact limits our quantitive analysis, the impact of the implanted muon on the spin density can still give us an insight into the magnitude of effects it could have on the magnetic properties of the system. 
In Fig.~\ref{sites2}, the displayed isosurfaces represent increases and decreases in spin density. The perturbation caused by the muon probe is quite dramatic: it results in the flipping of multiple spins.  In each case the spin on four Cu ions close to the muon sites are flipped (note the use of periodic boundary conditions).  These spin flips occur for Cu ions up to around 9~\AA\ away from the muon site, indicating a range over which muon-induced distortions to the electronic structure can occur.  While most of these changes in spin density are spin flips, the Cu atoms nearest the muon sites that dominate the local field at the muon site show a decrease in the magnitude of their spin.  For the  rung sites, Mulliken population analysis indicates a 15 \% reduction of the spin on the nearest Cu ion from 0.34$\hbar/2$ to 0.29$\hbar/2$.  A similar reduction of the Cu spin is seen for the  leg sites.  The tetrahedron sites see a 25 \% reduction of the nearest Cu moment, from around 0.33$\hbar/2$ to 0.25$\hbar/2$. This is likely to reflect the muon's closer proximity to a Cu ion for these latter sites.

\setlength{\tabcolsep}{2.2mm}
\begin{table*}
\begin{tabular}{| lcc | ccc | ccc | cccc |}
\hline
Site  & Energy & Number of  & \multicolumn{3}{c|}{FM$^a$} & \multicolumn{3}{c|} {AF$^b$} & \multicolumn{4}{c|}{Canted AF$^c$}\\
 & meV & structures & $|B_{\perp}|$ & $B_{\parallel}$ & $|B|$ & $|B_{\perp}|$ & $B_{\parallel}$ & $|B|$ & $|B_{\perp}|$ & $B_{\parallel}$ & $|B|$ & 0.41$B_{\parallel}$ \\ 
\hline 
Rung & 0 & 6  & 6 & 2.0 & 6.6 & 18 & $\pm$5.6 & 19 & 8.9, 15 & -2.1, 5.2 & 9.1, 16& 0.9, 2.1\\
Leg & 76 & 6  & 1.4 & -22 & 22 & 20 & $\pm$2.7 & 20 & 13, 14 & -19, -15 & 21, 23& -7.7, -6.3\\
Tetrahedron & 140 & 21 & 85 & -64 & 140 & 130 & $\pm$34 & 150 & 99, 130 & -71, -27 & 130, 150& -33, -11\\
\hline
\end{tabular}
\caption{Summary of the three muon sites and the average sensitivity of their local fields to different modes of correlated ordering of the Cu moments, expressed in units of mT/$(\mu_{\mathrm{Cu}}/\mu_{\mathrm{B}})$. Field directions $B_{\perp}$ and $B_{\parallel}$ are relative to the $b$ axis, which is the orientation of the applied field in the experiments.
Large $B_{\perp}$ values are associated with strong relaxation in the LF configuration and large $B_{\parallel}$ values are associated with strong relaxation and significant spectral shifts in the TF configuration.
$^a$FM moments along the $b$ axis. 
$^b$AF moments along the $c$ axis. 
$^c$Canted AF mode (Thielemann {\it et al.}  \cite{thielemann}), the final column lists the predicted field shifts in mT for the TF spectrum with the LRO moment of 0.41 $\mu_B$. 
\label{sitetable}}
\end{table*}

Since the muon has a large zero point energy due to its small mass, we expect quantum delocalisation across the closely spaced members of each group of sites, thus the properties of the effective quantum delocalised muon site defined by the group of structures can be estimated by taking an average over the group.
By exploring the effect of three characteristic modes of magnetic order on the dipolar field at the muon sites we can determine the relative sensitivity of the muon sites to different types of static order and fluctuations.  First we explore the effect of a a uniform FM order parameter along the $b$ axis,  then a staggered AF order parameter in the $ac$ plane directed along $c$.
Finally we take the known canted AF structure for the ordered state of the system, which allows an estimate the size of the local dipole field at the candidate muon sites in the LRO region.
The dipolar field experienced by a muon at position $\textbf{r}_{\mu}$ due to the magnetically ordered structure is given by
\begin{equation}
\textbf{B}_{\textrm{dipole}}(\textbf{r}_{\mu})=\sum_{i} \frac{\mu_0}{4 \pi r^3} \left[ 3(\boldsymbol{\mu}_i \cdot \hat{\textbf{r}}) \hat{\textbf{r}} - \boldsymbol{\mu}_i  \right],
\label{equation7}
\end{equation}
where $\mu_0$ is the permeability of free space and $\textbf{r}=\textbf{r}_{\mu}-\textbf{r}_i$ is the position of the muon relative to ion $i$ with magnetic moment $\boldsymbol{\mu}_i$.

Our dipolar field calculations are summarised in Table \ref{sitetable}.
The large transverse field fluctuation amplitude $B_{\perp}$ found for the tetrahedron site is expected to dominate the LF relaxation, with the greatest sensitivity being seen for AF or canted AF spin fluctuations. When considering the TF spectra this site also has the largest $B_{\parallel}$ coupling, giving a large negative shift for uniform moments polarised along the $b$ axis. The final column in Table \ref{sitetable} gives the TF spectral shift for the known LRO structure, which allows the satellite spectral features of Figures 2 to 4 to be clearly assigned. We see that the observed +2 mT feature corresponds to the rung site and the -6 mT feature corresponds to the leg site.

The sites found in the case of muonium (investigated by employing a neutral simulation cell) are very similar to those described above.  After relaxation of the structure, there is little electron density found around the muon, with the additional electron density instead moving to the piperidinium ion (close to the N) and the CuBr$_4$ unit closest to the muon.  The additional electron density around this Cu ion results in a slight reduction of the Cu moment compared with each of the corresponding cases for diamagnetic $\mu^+$.  The spin density around the muon is found to be small for both charged and neutral cells.  We therefore expect the contact hyperfine contribution to the local magnetic field experienced by the muon to be small, with dipolar coupling providing the dominant contribution.


\begin{thebibliography}{xx}
\bibitem{note1}
Amplitudes were allowed to vary slightly in the fitting procedure. There is a drop in the amplitude of the broad central feature when its width increases (rising again when the width decreases) consistent with a constant area of this feature in the frequency domain, and therefore for a constant contribution from muon sites. When the low field feature is no longer resolvable, it is not completely clear whether the spectral weight is lost from the total signal, owing to the relatively small size of this feature compared to the larger contributions to the central peak. However, the broadening out of this feature (and the loss of its spectral weight) would be consistent with the results of the fitting.
\end{thebibliography}
\end{document}